\documentclass[twocolumn,prl,showpacs,floatfix]{revtex4}

\usepackage{subfigure}
\usepackage{graphicx}

\begin{document}

%\begin{document}
%\draft 
%\preprint{UFIFT-HEP-11-1}
%\date{\today}

\title{Axion Dark Matter and Cosmological Parameters}

\author{O. Erken, P. Sikivie, H. Tam, Q. Yang}

\affiliation{Department of Physics, University of Florida, 
Gainesville, FL 32611, USA}

\begin{abstract}

We observe that photon cooling after big bang nucleosynthesis (BBN) but
before recombination can remove the conflict between the observed and
theoretically predicted value of the primordial abundance of $^7$Li.  
Such cooling is ordinarily difficult to achieve.  However, the recent
realization that dark matter axions form a Bose-Einstein condensate 
(BEC) provides a possible mechanism, because the much colder axions 
may reach thermal contact with the photons.  This proposal predicts 
a high effective number of neutrinos as measured by the cosmic 
microwave anisotropy spectrum.

\end{abstract}

\maketitle

\section{Introduction}   

The agreement between observations and the BBN predictions for the 
primordial abundances of light elements is often touted as a triumph 
of the standard $\Lambda$CDM cosmological model. Under the assumption 
that there are three neutrino species, BBN as a theory requires essentially 
a single input: the baryon-to-photon ratio, usually given by the parameter 
$\eta_{10} = 10^{10}n_B/n_{\gamma}$ \cite{Cyburt:2008kw}. If one takes 
$\eta_{10}$ to be $6.190\pm0.145$, in accordance with the latest Wilkinson 
Microwave Anisotropy Probe(WMAP) results \cite{Komatsu:2010fb,Steigman:2010zz}, 
the inferred primordial abundances of the majority of the light elements 
(D, $^4$He, $^3$He) are remarkably consistent with BBN predictions, save 
one exception: that of $^7$Li is approximately two to three times less 
than what the theory predicts. The discrepancy is deemed statistically 
significant, and there is so far no widely accepted explanation for the 
anomaly. In the literature, this is referred to as the ``Lithium Problem''.
%"commonly referred to..."

One of the most difficult issues involved in testing BBN is how reliably 
to infer the primordial abundances of light elements from measurements 
that are available to us.  Subsequent to BBN, the original relic abundances 
are all subject to further modification by complicated stellar processes.  
$^7$Li, for example, can be both depleted and synthesized in stars, as well 
as produced by cosmic-ray nucleosynthesis.  As such, the abundance of $^7$Li 
is inferred primarily from absorption lines in the atmosphere of galactic 
halo stars with low metallicity, since these stars are very old and have 
experienced very little nuclear processing (See \cite{Steigman:2005uz,
Steigman:2007xt, Cyburt:2008kw,Steigman:2010zz} for details).

%(Excellent reviews on post-BBN
%$^7$Li evolution can be found in refs. \cite{Steigman:2005uz,Steigman:2007xt,
%Cyburt:2008kw,Steigman:2010zz}.)

Although these post-BBN effects lead to considerable complication, they also open
up many different avenues to explain the $^7$Li anomaly.  For many years, it has
been hoped that better determination of nuclear parameters will gradually narrow
the discrepancy, though it was eventually realized that does not seem achievable
\cite{hope}. Quite the contrary, it was found in \cite{Cyburt:2008kw} that
improved data on the neutron life-time and the cross sections p(n,$\gamma$)d
and $^3$He($\alpha,\gamma$)$^7$Be increases the predicted abundance of $^7$Li,
worsening the disagreement.  Revisions to stellar evolution, as a consequence of
systematic errors in the effective temperature of the metal-poor stars
\cite{Melendez:2004ni, Fields:2004ug}, and surface $^7$Li depletion
in the interior of stars due to some mixing or diffusive processes
\cite{systerr}, have also been investigated as possible solutions,
but are still considered controversial \cite{Cyburt:2008kw}.

The fact that the nuclear reactions relevant to the production of   
both primordial and post-BBN $^7$Li are now quite well understood has led to
speculations that the anomaly might instead be caused by new physics.  Many  
explanations have been proposed, such as the variation in time of the deuteron
binding energy and of fundamental couplings \cite{Dmitriev:2003qq, Coc:2006sx},
and the decay of a relatively long-lived particle in the context of supersymmetry
\cite{susy}. At this point, none of these explanations have won general acceptance
in the cosmology community.

In this paper, we propose a way to remove the conflict between data and theory
for the abundance of $^7$Li, based on the cooling of photons between the end of
BBN and decoupling.  Processes that do this are difficult to come by.  Indeed,
typical processes arising from new physics tend to heat up the photons, modifying 
$\eta_{10}$ in the wrong direction \cite{Cyburt:2008kw}.  However, the recent
realization that dark matter axions form a BEC at approximately 500 eV photon 
temperature \cite{CABEC} provides a possible mechanism \cite{therm}.  Essentially, 
the high occupation of axion modes with very low momenta greatly enhances the 
strength of their gravitational interactions, such that an exchange of energy 
between the photons and the much colder axions becomes possible.  Photon cooling 
implies that $\eta_{10,{\rm BBN}}< \eta_{10,{\rm WMAP}}$, which has the effect 
of  reducing the production of $^7$Li \cite{Steigman:2007xt}.  If thermal
equilibrium between the photons and axions is achieved, the $^7$Li abundance 
is reduced by approximately a factor 2 (see below), alleviating the
discrepancy and perhaps removing it altogether. However, our proposal 
predicts a higher abundance for D than present observations indicate 
and predicts that the {\it effective} number of thermally excited 
neutrino degrees of freedom is high: $N_{\rm eff}$ = 6.77.

Photon cooling by kinetic mixing with hidden photons was proposed in 
ref.~\cite{kinmix}.

\section{Dark matter axions}

The axion was originally postulated to explain the absence of CP violation   
in the strong interactions \cite{axion}. It was later realized that the
population of axions produced by the turn on of the axion mass during the
QCD phase transition possesses the right properties to be cold dark matter
(CDM) \cite{axdm}.  First, they have the measured CDM  density if the axion
mass $m$ is of order $10^{-5} {\rm eV}/c^2$ \cite{axden}.  
%(For a pedagogical review on cold axion cosmological abundance,  see 
%\cite{Sikivie:2006ni}.  
Second, the axions thus produced are
very cold: their average momentum is only of order the Hubble expansion
rate ($3\times 10^{-9} {\rm eV}/\hbar$) when the axion mass effectively
turns on and has been redshifting ever since to a mere $10^{-17}mc$ today. 
Third, axions in this mass range interact very weakly through all forces
other than gravity.  These properties make axions one of the leading
candidates for CDM.  The other main contenders are weakly interacting
massive particles (WIMPs) and sterile neutrinos.  Henceforth, we set 
$\hbar=c=1$.

Observationally it seems difficult to distinguish among the CDM candidates.
However, it was recently realized that axions form a BEC
through gravitational self-interactions at approximately a photon temperature of
500 eV \cite{CABEC,therm}.  The relaxation rate of cold dark  
matter axions through gravitational self-interactions is of order
\begin{equation}
\Gamma_a \sim 4 \pi G n m^2 \ell^2
\label{arelax}
\end{equation}
where $n$ is the number density of cold axions and
$\ell \sim t_1 {a(t) \over a(t_1)}$ their correlation length.
$t_1 \sim 2 \cdot 10^{-7}$ s is the time at which the axion mass
effectively turns on during the QCD phase transition, and $a(t)$ 
is the scale factor.  Eq.~(\ref{arelax}) is appropriate when the
energy dispersion of the particles is less than their relaxation rate.
We refer to this case as `the condensed regime', to distinguish it from
the more commonly encountered `particle kinetic regime' defined by the
condition that the energy dispersion of the particles is large compared
to the relaxation rate.  Cold dark matter axions are in the condensed
regime after $t_1$.  The ratio $\Gamma_a(t)/H(t)$, where $H(t)$ is the
Hubble rate, is of order $5 \cdot 10^{-7}$ at $t_1$ but increases
with time as $a(t)^{-1} t$, and reaches one at approximately 500 eV photon
temperature.  At that time the axions thermalize and form a BEC.  Almost
all axions go to the lowest energy state available.  The correlation
length grows and becomes of order the horizon.  In the linear regime
of evolution of density perturbations and within the horizon, the lowest
energy state is time independent and no rethermalization is necessary for
the axions to remain in the lowest energy state.  In that case, axion BEC
and ordinary CDM are indistinguishable on all scales of observational
interest \cite{CABEC}.  However, beyond first order perturbation theory
and/or upon entering the horizon, the axions rethermalize to try and remain
in the lowest energy available state.  Axion BEC behaves differently from
CDM then and the resulting differences are observable.

The study of the catastrophe structure of the inner caustics of galactic halos
provides evidence that the dark matter is an axion BEC \cite{CABEC,case}. Briefly, 
there is a dichotomy in the classification of the inner caustics in terms of their
catastrophe structure, depending on the angular momentum distribution of the infalling
particles \cite{inner}. Axions in a BEC are in a state of net overall rotation and 
produce caustic rings, whereas ordinary CDM has an irrotational velocity field and 
produces tent-like caustics. There are several pieces of evidence for the existence 
of caustic rings at the predicted radii in various galaxies \cite{selfsim, Duffy}. 
It is shown in ref. \cite{case} that the phase space structure implied by the 
evidence for caustic rings is precisely and in all respects that predicted by 
the assumption that dark matter is a rethermalizing axion BEC.

\section{Photon Cooling}

The Lithium Problem refers to the mismatch between the observed and predicted
abundance of primordial $^7$Li by a factor 2 or 3 \cite{Steigman:2005uz, 
Cyburt:2008kw,Steigman:2010zz}.  For $\eta_{10} \gtrsim 2.7$, the predicted 
$^7$Li abundance increases with $\eta_{10}$.  Hence, a cooling of the photons 
between the end of BBN and decoupling reduces the discrepancy.

The gravitational fields of the cold axion fluid cause transitions between
momentum states of other particle species present.  For particles which are 
bosons or non degenerate fermions, the relaxation rate through gravitational
interactions with the cold axions is of order \cite{therm}
\begin{equation}
\Gamma \sim 4 \pi G m n \ell {\omega \over \Delta p}  
\label{general}
\end{equation}
where $\omega$ is the typical energy of the particles and $\Delta p$ their
momentum dispersion.  Eq.~(\ref{general}) generalizes Eq.~(\ref{arelax}) to
other species that are in the presence of the cold axions.  [Eq.~(\ref{arelax})
follows from Eq.~(\ref{general}) by setting $\omega = m$ and $\Delta p = \ell^{-1}$ 
as is appropriate for the cold axions themselves.]  For photons to cool substantially
it is necessary that energy is transferred from the photons to the low momentum 
highly occupied axion states and from those to the relativistic axion states.  For 
both relativistic axion states and for photons, $\Delta p \sim \omega$ and hence 
their relaxation rate $\Gamma_r$ through gravitational interactions with cold 
axions is of order $4 \pi G n m \ell$.  Using the Friedmann equation, one finds 
that $\Gamma_r/H \propto a(t)$ before equality between matter and radiation and
remains constant after that.  At equality, $\Gamma_r/H|_{t_{\rm eq}} \sim 
\ell(t_{\rm eq})/t_{\rm eq}$.  If $\ell/t$ is order one at equality, the 
photons reach thermal equilibrium with the axions and hence cool.

Gravitational interactions conserve particle number and therefore produce 
only kinetic (as opposed to chemical) equilibrium between the species involved.
Also, after 500 eV photon temperature, the coupling between photons and baryons 
is in the kinetic, rather than chemical, equilibrium regime \cite{Hu:1995em}.   
Upon cooling, the photons that cannot be accommodated in thermally excited 
states enter the ground state, a plasma oscillation with zero wavevector.  
Since the photon chemical potential remains zero, the final photon spectrum 
is Planckian, consistent with observation.

Eq.~(\ref{general}) does not apply to degenerate fermions because of 
Pauli blocking.  The cosmic neutrinos are semi-degenerate since they 
have a thermal distribution with zero chemical potential.  Their 
thermalization rate is less than that $\Gamma_r$ of relativistic 
bosons.  Since $\Gamma_r/H \propto t n \ell \propto t^2 a^{-3}(t)$,
that ratio does not grow after equality.  Since the relativistic axions 
may only reach thermal contact with the cold axions at equality and the 
neutrinos are delayed relative to the relativistic axions, we believe it 
most likely that neutrinos remain decoupled from the axions, photons and 
baryons at all times.

It is straightforward to determine how much the photons cool if they 
reach thermal equilibrium  with the axions.  Energy conservation implies
$\rho_{i,\gamma} = \rho_{f,\gamma} + \rho_{f,a}$ because the contributions
to the energy density of the initial axions and of the baryons are negligible.
The ratio between the final and initial photon temperature is thus $(2/3)^{1/4}.$
Since their number density is proportional to $T^3$, we find:
\begin{equation}
\eta_{10,{\rm BBN}} = \left(\frac{2}{3}\right)^{3/4}\eta_{10,{\rm WMAP}}
= 4.57 \pm 0.11
\end{equation}
using $\eta_{10,WMAP} = 6.190\pm 0.145$ \cite{Komatsu:2010fb}.  Because 
the $^7$Li abundance is proportional to $\eta_{10,{\rm BBN}}^2$ in the 
range of interest, it is reduced by approximately the factor
$({2 \over 3})^{3 \over 2} \simeq$ 0.55.

A number of authors proposed earlier that the dark matter is a BEC 
\cite{otherbec}.  The photon cooling described here may occur in those 
cases as well.  If the particles are in the condensed regime, the 
relaxation rate is given by Eq.~(\ref{arelax}), calculated with 
appropriate values for $n$, $m$, and $\ell$.  However, in many of 
these proposals, the cosmological history of the dark matter particle 
is not known, rendering the computation of the relaxation rate difficult.

\section{Cosmological consequences}
\subsection{Effect on light element primordial abundances}

Whether photon cooling by axion BEC solves the Lithium Problem remains
to be seen.  The data have been time dependent in addition to the usual  
uncertainties.  In Fig. 1, we plot the value of $\eta_{10, \rm{BBN}}$
in the standard cosmological model, labeled `WIMP', and in the scenario
described here, labeled `axion', along with the values inferred from the    
observed light element abundances according to the review by G. Steigman
in 2005 \cite{Steigman:2005uz}, the review by F. Iocco et al. in 2008
\cite{Iocco} and a private communication from G. Steigman updating his
2005 estimates in the light of recent observations \cite{Steigman_email}.      
The error bars indicate the range of $\eta_{10, \rm{BBN}}$ consistent with
the estimated 1-$\sigma$ uncertainties in the observations.  The axion
prediction agrees very well with the $^7$Li abundance at the time
of Steigman's 2005 review ($\eta_{10,^7{\rm Li}} = 4.50 \pm 0.30$).
However more recent observations indicate a lower primordial $^7$Li   
abundance, worsening the Lithium Problem.

Perhaps more problematic is that a smaller $\eta_{10,{\rm BBN}}$ predicts an
overproduction of D.  Traditionally, D has been the prime choice as a baryometer
among the light elements, due to its sensitivity to $\eta_{10,{\rm BBN}}$ and
simple post-BBN evolution (abundance monotonically decreasing).  The major
drawback with D is that its abundance is inferred from a very small set of
(seven) spectra of QSO absorption line systems \cite{Pettini:2008}.  Worse
yet, these few measurements have a large dispersion, and do not seem to
correlate with metallicity, obscuring the expected deuterium plateau.
Due to the various inadequacies in the D measurements mentioned, we have
reservations about the common practice of attaching most significance on
D in the comparison between data and BBN predictions.  In comparison, $^7$Li
is inferred from a large number of measurements, which are more-or-less
consistent. Also, since D is more easily destructible than $^7$Li, it is
conceivable that unknown stellar processes further deplete D.

Finally the $^3$He and $^4$He inferred $\eta_{10,{\rm BBN}}$ values have
large error bars and hence carry less statistical weight.  The $^4$He
inferred value has increased recently ($5.5 < \eta_{10,^4{\rm He}} < 11$
according to ref. \cite{Iocco} and $7.5 < \eta_{10,^4{\rm He}} < 20$
according to ref. \cite{Steigman_email}) compared to its accepted
value a few years ago, creating additional uncertainty.

\begin{figure} 
\includegraphics[width=1.0\columnwidth]{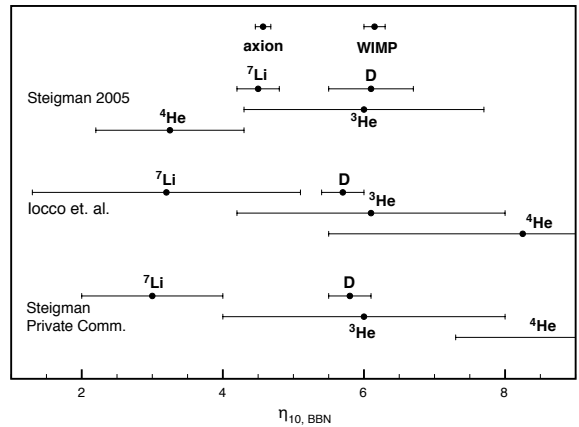}
\caption{Values of $\eta_{10,{\rm BBN}}$ inferred from the abundances
of $^7$Li, D, $^3$He and $^4$He, and the predicted values in the standard  
cosmological model (WIMP) and in our proposal (axion).  The data inferred
values are taken from refs. \cite{Steigman:2005uz}, \cite{Iocco} and
\cite{Steigman_email}.  The error bars indicate the $\eta_{10,{\rm BBN}}$
values consistent with the estimated 1-$\sigma$ uncertainties in the
observations.}
\label{sig}
\end{figure}

\subsection{Effective number of neutrino species}

After the axions are heated up and reach the same temperature as the
photons, most of them are still in the ground state.  The axions in the 
ground state behave as cold dark matter.  The axions in the excited states
contribute one bosonic degree of freedom to radiation.  The radiation 
content of the universe is commonly given in terms of the effective
number $N_{\rm eff}$ of thermally excited neutrino degrees of freedom,
defined by
\begin{equation}
\rho_{\rm rad} = \rho_\gamma [1 +
N_{\rm eff} {7 \over 8} \left({4 \over 11}\right)^{4 \over 3}]
\label{Neff}
\end{equation}
where $\rho_{\rm rad}$ is the total energy density in radiation and 
$\rho_\gamma$ is the energy density in photons only.  The standard 
cosmological model with ordinary cold dark matter predicts $N_{\rm eff}$ 
= 3.046, slightly larger than 3 because the three neutrinos heat up a 
little during $e^+ e^-$ annihilation.  Taking account of the fact that 
not only is there an extra species of radiation (thermally excited axions) 
but also the contribution of the three ordinary neutrinos is boosted because 
the photons have been cooled relative to them, the proposed scenario predicts
\begin{eqnarray}
\rho_{\rm rad} &=& \rho_\gamma + \rho_a + \rho_\nu \nonumber\\
&=& \rho_\gamma \left[1+\frac{1}{2} + 
3.046~{7 \over 8} \left({4 \over 11}\right)^{4 \over 3}{3 \over 2} \right],
\end{eqnarray}
which yields $N_{\rm eff}$ = 6.77.

At present, the measured values are smaller than this prediction. The WMAP
collaboration found $N_{\rm eff} = 4.34_{-0.88}^{+0.86}$(68\% CL) based on 
their 7 year data combined with independent data on large scale structure 
and the Hubble constant \cite{Komatsu:2010fb}.  An analysis \cite{Hamann} 
using the Sloan Digital Sky Survey (SDSS) data release 7 halo power spectrum 
found $N_{\rm eff} = 4.8 \pm 2.0$ (95\% CL). The Atacama Cosmology Telescope
(ACT) collaboration finds \cite{Dunkley:2010ge} $N_{\rm eff} = 5.3\pm 1.3$ 
(68\%CL) using only their CMB anisotropy data and $N_{\rm eff} = 4.56 \pm 0.75$ 
(68\% CL) when combining that data with large scale structure data.  The 
tendency for the measured values to be larger than 3.046 has been taken 
sufficiently seriously to prompt proposals for new physics involving 
extra neutrino species or a neutrino asymmetry \cite{exneu}.  The Planck 
mission is expected to measure $N_{\rm eff}$ with much greater precision 
\cite{Ichi}.  In so doing, it may shed light on the nature of dark matter.

We thank Gary Steigman for extended discussions, and our referees for 
numerous suggestions for improvement.  This work was supported in part 
by the U.S. DOE under contract DE-FG02-97ER41029.


\begin{thebibliography}{}

\bibitem{Cyburt:2008kw}
  R.~H.~Cyburt, B.~D.~Fields, K.~A.~Olive,
  %``A Bitter Pill: The Primordial Lithium Problem Worsens,''
  JCAP {\bf 0811}, 012 (2008).

\bibitem{Komatsu:2010fb}
  E.~Komatsu {\it et al.},
  %``Seven-Year Wilkinson Microwave Anisotropy Probe (WMAP) Observations: 
  %Cosmological Interpretation,''
  Ap.J. Suppl. 192:18 (2011).
  %%CITATION = ARXIV:1001.4538;%%

\bibitem{Steigman:2010zz}
  G.~Steigman,
  %``Primordial Nucleosynthesis: The Predicted and Observed Abundances and Their
  %Consequences,''
  arXiv:1008.4765.

\bibitem{Steigman:2005uz}
  G.~Steigman,
  %``Primordial Nucleosynthesis: Successes And Challenges,''  
  Int.\ J.\ Mod.\ Phys.\  E {\bf 15}, 1 (2006).

\bibitem{Steigman:2007xt}
  G.~Steigman,
  %``Primordial Nucleosynthesis in the Precision Cosmology Era,''
  Ann.\ Rev.\ Nucl.\ Part.\ Sci.\  {\bf 57}, 463 (2007).

\bibitem{hope}
  A.~Coc {\it et al.},
   %``Updated Big Bang Nucleosynthesis confronted to WMAP observations and to the
   %Abundance of Light Elements,''
   Ap. J.  {\bf 600}, 544 (2004);
  R.~H.~Cyburt, B.~D.~Fields, K.~A.~Olive,
   %``Solar Neutrino Constraints on the BBN Production of Li,''
   Phys.\ Rev.\  D {\bf 69}, 123519 (2004);
  C.~Angulo {\it et al.},
   %``The 7Be(d,p)2alpha cross section at Big Bang energies and the primordial
   %7Li abundance,''
   Ap. J.  {\bf 630}, L105 (2005).

\bibitem{Melendez:2004ni}
  J.~Melendez, I.~Ramirez,
  %``Reappraising the Spite lithium plateau: Extremely thin and marginally
  %consistent with WMAP,''
  Ap. J.  {\bf 615}, L33 (2004).

\bibitem{Fields:2004ug}
  B.~D.~Fields, K.~A.~Olive, E.~Vangioni-Flam,
  %``Implications of a new temperature scale for halo dwarfs on LiBeB and
  %chemical evolution,''
  Ap. J.  {\bf 623}, 1083-1091 (2005).

\bibitem{systerr}
  S.~Vauclair, C.~Charbonnel,
   %``Element segregation in low metallicity stars and the primordial lithium
   %abundance,''
   Ap. J. {\bf 502} 372 (1998);
  M.~H.~Pinsonneault {\it et al.},
   %``Halo star lithium depletion,''
   Ap. J. {\bf 527}, 180-198 (2002);
  M.~H.~Pinsonneault {\it et al.},
   %``Stellar mixing and the primordial lithium abundance,''  
   Ap. J. {\bf 574}, 398-411 (2002); 
  O.~Richard, G.~Michaud, J.~Richer,
   %``Implications of WMAP observations on Li abundance and stellar evolution
   %models,'' 
  Ap. J.  {\bf 619}, 538-548 (2005);
  A.J. Korn et al., Ap. J. {\bf 442} (2006) 657.
  
\bibitem{Dmitriev:2003qq}
  V.~F.~Dmitriev, V.~V.~Flambaum, J.~K.~Webb,
  %``Cosmological variation of deuteron binding energy, strong interaction  and
  %quark masses from big bang nucleosynthesis,''
  Phys.\ Rev.\  D {\bf 69}, 063506 (2004).
  
\bibitem{Coc:2006sx}
  A.~Coc {\it et al.},
  %``Coupled Variations of Fundamental Couplings and Primordial
  %Nucleosynthesis,''
  Phys.\ Rev.\  D {\bf 76}, 023511 (2007).
   
\bibitem{susy}
  K.~Jedamzik,
   %``Did something decay, evaporate, or annihilate during big bang
   %nucleosynthesis?,''  
   Phys.\ Rev.\  D {\bf 70}, 063524 (2004);
  J.~L.~Feng, S.~Su, F.~Takayama,
   %``Supergravity with a gravitino LSP,'' 
   Phys.\ Rev.\  D {\bf 70}, 075019 (2004);
  J.~R.~Ellis, K.~A.~Olive, E.~Vangioni,
   %``Effects of unstable particles on light-element abundances: Lithium  versus
   %deuterium and He-3,''
   Phys.\ Lett.\  B {\bf 619} (2005) 30;
  K.~Jedamzik {\it et al.},
   %``Solving the cosmic lithium problems with gravitino dark matter in the
   %CMSSM,''
   JCAP {\bf 0607}, 007 (2006);
  R.~H.~Cyburt {\it et al.},
   %``Bound-state effects on light-element abundances in gravitino dark  matter
   %scenarios,''
   JCAP {\bf 0611} (2006) 014;
  T.~Jittoh {\it et al.}, 
   %``Possible solution to the Li-7 problem by the long lived stau,''
   Phys. Rev. {\bf D76}, 125023 (2007);
  K.~Jedamzik, M.~Pospelov,
   %``Big Bang Nucleosynthesis and Particle Dark Matter,''
   NJP  {\bf 11}, 105028 (2009).
   %New J. Phys.
   
\bibitem{CABEC}
  P.~Sikivie, Q.~Yang,
  %``Bose-Einstein Condensation of Dark Matter Axions,''
  Phys.\ Rev.\ Lett.\  {\bf 103}, 111301 (2009).
  %[arXiv:0901.1106 [hep-ph]].
   
\bibitem{therm}
  O. Erken, P. Sikivie, H. Tam, Q. Yang, arXiv:1111.1157.

\bibitem{kinmix}
  J.~Jaeckel, J.~Redondo, A.~Ringwald, 
   %``Signatures of a hidden cosmic microwave background,''
   Phys. Rev. Lett. {\bf 101} 131801 (2008).

\bibitem{axion}
  R.~D.~Peccei, H.~R.~Quinn,
   %``CP Conservation in the Presence of Instantons,''
   Phys.\ Rev.\ Lett.\  {\bf 38}, 1440-1443 (1977), and
 %R.~D.~Peccei, H.~R.~Quinn,
   %``Constraints Imposed by CP Conservation in the Presence of Instantons,''  
   Phys.\ Rev.\  {\bf D16}, 1791-1797 (1977);   
  S.~Weinberg,
   %``A New Light Boson?,''
   Phys.\ Rev.\ Lett.\  {\bf 40}, 223-226 (1978);
  F.~Wilczek,
   %``Problem of Strong p and t Invariance in the Presence of Instantons,''
   Phys.\ Rev.\ Lett.\  {\bf 40}, 279-282 (1978).
  
\bibitem{axdm}
J. Preskill, M. Wise, F. Wilczek, Phys. Lett. {\bf B120} (1983) 127;
L. Abbott, P. Sikivie, Phys. Lett. {\bf B120} (1983) 133;
M. Dine, W. Fischler, Phys. Lett. {\bf B120} (1983) 137;
%J. Ipser and P. Sikivie, Phys. Rev. Lett. {\bf 50} (1983) 925.

\bibitem{axden}
  For a review see P.~Sikivie, Lect.\ Notes Phys.\  {\bf 741}, 19-50 (2008).  More 
  recent papers on axion radiation by strings include: 
  O.~Wantz, E.~P.~S.~Shellard, Phys.\ Rev.\  {\bf D82}, 123508 (2010);
 %T.~Hiramatsu, M.~Kawasaki, T.~Sekiguchi, M.~Yamaguchi, J.~'i.~Yokoyama,  
 %[arXiv:1012.5502 [hep-ph]].
  T.~Hiramatsu {\it et al.}, arXiv:1012.5502.
     
\bibitem{case}
  P.~Sikivie,
  %``The emerging case for axion dark matter,''
  Phys.\ Lett.\  B {\bf 695}, 22 (2011).   
   
\bibitem{inner}
  A.~Natarajan, P.~Sikivie,
  %``The inner caustics of cold dark matter halos,''
  Phys.\ Rev.\  {\bf D73}, 023510 (2006).
   
\bibitem{selfsim}
  J.A. Fillmore, P. Goldreich, Ap. J. {\bf 281}, 1 (1984);
  E. Bertschinger, Ap. J. Suppl. {\bf 58}, 39 (1985);
  P. Sikivie, I. Tkachev, Y. Wang, Phys. Rev. Lett. {\bf 75}, 2911 (1995)
  and Phys. Rev. D56 (1997) 1863.
  
\bibitem{Duffy}
  L.~D.~Duffy, P.~Sikivie,
  %``The Caustic Ring Model of the Milky Way Halo,''
  Phys.\ Rev.\  {\bf D78}, 063508 (2008).
  
%\cite{Hu:1995em}
\bibitem{Hu:1995em}
  W.~T.~Hu, Ph.D. thesis,
  %``Wandering in the Background: A CMB Explorer,''
  astro-ph/9508126 and references therein.
  
\bibitem{otherbec}
  S.~-J.~Sin, Phys.\ Rev.\  {\bf D50}, 3650-3654 (1994);
  W.~Hu, R.~Barkana, A.~Gruzinov, Phys.\ Rev.\ Lett.\  {\bf 85}, 1158-1161 (2000);
  J.~-W.~Lee, S.~Lim, JCAP {\bf 1001}, 007 (2010);
  E.~W.~Mielke, J.~A.~V.~Perez, Phys.\ Lett.\  {\bf B671}, 174-178 (2009);
  F.~Ferrer, J.~A.~Grifols, JCAP {\bf 0412}, 012 (2004);
.  C.~G.~Boehmer, T.~Harko, JCAP {\bf 0706}, 025 (2007).
  
\bibitem{Iocco}
F. Iocco et al.,
Phys. Rep. {\bf 472} (2009) 1.
   
\bibitem{Steigman_email}
  G. Steigman, private correspondence.
   
\bibitem{Pettini:2008}
  M. Pettini {\it et al.}, MNRAS, {\bf 391} (2008) 1499.
   
\bibitem{Hamann}
  J. Hamann {\it et al.},   
  JCAP {\bf 07} (2010) 022.
   
\bibitem{Dunkley:2010ge}
  J.~Dunkley {\it et al.}, 
  %``The Atacama Cosmology Telescope: Cosmological Parameters from the 2008
  %Power Spectra,''
  arXiv:1009.0866.
   
\bibitem{exneu}
  L.M. Krauss, C. Lunardini, C. Smith, arXiv:1009.4666;
  J. Hamann {\it et al.}, Phys. Rev. Lett. {\bf 105} (2010) 181301.

\bibitem{Ichi}
  K. Ichikawa, T. Sekiguchi, T. Takahashi,
  Phys. Rev. {\bf D78} (2008) 083526.

\end{thebibliography}
\end{document}